# ON THE BI-HAMILTONIAN THEORY FOR THE HARRY DYM EQUATION

MARCO PEDRONI, VINCENZO SCIACCA, AND JORGE P. ZUBELLI

ABSTRACT. We describe how the Harry Dym equation fits into the the bi-Hamiltonian formalism for the Korteweg-de Vries equation and other soliton equations. This is achieved by means of a certain Poisson pencil constructed from two compatible Poisson structures. We obtain an analogue of the Kadomtsev-Petviashivili hierarchy whose reduction leads to the Harry Dym hierarchy. We call such system the HD-KP hierarchy. Then, we construct an infinite system of ordinary differential equations (in infinitely many variables) that is equivalent to the HD-KP hierarchy. Its role is analogous to the one played by the Central System for the Kadomtsev-Petviashivili hierarchy.

## 1. INTRODUCTION

The ubiquitous presence of completely integrable models, such as the nonlinear-Schrödinger equation, the Korteweg-de Vries equation, and the $N$-wave system, in a wide variety of physical phenomena explains part of the attention soliton equations have received during the last thirty years. Another reason for such attention is the multitude of deep connections of soliton equations with different branches of mathematics. They range from algebraic geometry to numerical analysis and include operator theory as well as symplectic geometry.

Directly connected with the Korteweg-de Vries (KdV) equation one can find an intriguing equation known as the Harry Dym (HD) equation

(1.1) $$q_t = 2(1/\sqrt{(1+q)})_{xxx}$$

or equivalently

(1.2) $$\rho_t = \rho^3 \rho_{xxx}$$

after the substitution $\rho = -(1+q)^{-1/2}$.

Equation (1.1) was discovered in an unpublished work by Harry Dym [1], and appeared in a more general form in works of P. C. Sabatier [2, 3, 4]. It possesses many of the typical properties of soliton equations. See [5, 6, 7, 8, 9, 10] and references therein. More recently, its relations with the Kadomtsev-Petviashvili (KP) and modified-KP hierarchy have been studied in detail by Oevel and Carillo [11].

Research at MSRI is supported in part by NSF grant DMS-9810361. M. Pedroni was supported by MURST under grant "Geometry of Integrable Systems." V. Sciacca was supported by MURST under grant "Non linear Mathematical Problems of Wave Propagation and Stability in Models of Continuous Media." J. P. Zubelli was supported by the fellowship from the Brazilian National Science Research Council (CNPq) of number 200258/84-2 ("Bolsista do CNPq - Brasil"), a grant from the State of Rio de Janeiro Research Foundation (FAPERJ) number 170.369/2001.

In the present work we discuss the HD equation from the bi-Hamiltonian point of view and show that it is amenable to the systematic treatment developed in [12, 13, 14]. More precisely, we start from a pair of compatible Poisson structures, which also play an important role in the KdV hierarchy, and construct a Casimir for the corresponding Poisson pencil. By virtue of the standard construction the full hierarchy of Hamiltonian flows commuting with HD emerges. We then construct, in an elementary way and without resorting to pseudo-differential operators, a hierarchy of commuting flows which by reduction leads to the HD hierarchy. This reduction process is totally analogous to the one that leads from the KP hierarchy to the KdV hierarchy, which justifies our calling such hierarchy the HD-KP hierarchy [1]. Finally, we write down a system of infinitely many ordinary differential equations that admits the HD-KP hierarchy as a suitable reduction.

The plan for the present work is the following:

In Section 1.1 we review the general definitions of Poisson geometry, the bi-Hamiltonian theory, and the construction of the KdV and KP hierarchies from this view point. This will serve as a blue-print for the HD theory developed later in this work and will also help us emphasize the parallels between the HD-KP hierarchy and the usual KP hierarchy.

Section 2 is devoted to exhibiting the bi-Hamiltonian structure of the HD equation and to constructing an infinite number of commuting flows from a solution of a Riccati equation. They will constitute the HD hierarchy.

Section 3 describes the construction of the HD-KP hierarchy, i.e., a hierarchy that plays the same role for HD as the KP plays for the KdV. In our bi-Hamiltonian formalism, this hierarchy is depicted as an infinite number of conservation laws. The currents for such conservation laws are studied in Section 4. In that section we also show that we can use such currents to fully describe the evolution of the quantities in the HD-KP hierarchy as ordinary differential equations. This leads to the notion of the *Central System*, which was also successfully used in earlier works dealing with the KdV and hidden-KdV hierarchies [12, 13, 16, 17].

1.1. **Preliminaries: The bi-Hamiltonian Approach to Soliton Equations.** In order to make the present work as much self-contained as possible we shall present in this sub-section an overview of the bi-Hamiltonian approach to the KdV and KP hierarchies. For a more detailed view and motivations the reader is referred to [12, 13, 14, 15, 16, 17, 18, 19].

Our approach is based on the concept of Poisson manifolds, which is the natural setting to study Hamiltonian systems.

**Definition 1.1.** *Let $\mathcal{M}$ be a differentiable manifold. A* Poisson bracket *on $\mathcal{M}$ is an anti-symmetric $\mathbb{R}$–bilinear map*

$$(1.3) \qquad \{\cdot,\cdot\} : C^\infty(\mathcal{M}) \times C^\infty(\mathcal{M}) \to C^\infty(\mathcal{M})$$

*with the following properties:*

---

[1] In fact, Carillo and Oevel construct in [11] such HD-KP hierarchy making use of pseudo-differential operators.

1. *Jacobi identity:* $\{\{F,G\},H\} + \{\{H,F\},G\} + \{\{G,H\},F\} = 0$.
2. *Leibniz rule*

$$\{FG, H\} = F\{G, H\} + \{F, H\}G.$$

*A* Poisson manifold *is a differentiable manifold endowed with a Poisson bracket.*

Starting from a Poisson bracket one can introduce a tensor $P$ of type (2,0), which we shall consider as a map from $T^*\mathcal{M}$ to $T\mathcal{M}$, defined by

(1.4) $$\langle dG, PdF \rangle = \{F, G\}.$$

This tensor is called the *Poisson tensor* associated with $\{\cdot,\cdot\}$. In a Poisson manifold, the vector field $X_H = \{H,\cdot\} = PdH$ is called the *Hamiltonian vector field associated with $H$*. The map $H \mapsto X_H$, assigning to a function $H$ its Hamiltonian vector field $X_H$, is a Lie algebra homomorphism:

(1.5) $$X_{\{F,G\}} = [X_F, X_G].$$

It follows that $\{F, G\} = 0$ implies $[X_F, X_G] = 0$; in other words, Hamiltonian vector fields associated with functions in involution commute. A *Casimir function* is a function $H$ such that $X_H = 0$, i.e., a function that is in involution with any other function on $\mathcal{M}$. In terms of the Poisson tensor, a Casimir function is a function whose differential belongs to the kernel of $P$: $PdH = 0$.

**Definition 1.2.** *Let $\mathcal{M}$ be a manifold endowed with two Poisson brackets, $\{\cdot,\cdot\}_0$ and $\{\cdot,\cdot\}_1$. These brackets are said to be* compatible *if any linear combination*

(1.6) $$\{F,G\}_\lambda \stackrel{def}{=} \{F,G\}_1 - \lambda\{F,G\}_0$$

*is still a Poisson bracket. A* bi-Hamiltonian manifold *is a manifold $\mathcal{M}$ endowed with two compatible Poisson brackets. In this case the bracket $\{\cdot,\cdot\}_\lambda$ is called the* Poisson pencil *defined by $\{\cdot,\cdot\}_0$ and $\{\cdot,\cdot\}_1$ on $\mathcal{M}$.*

Since antisymmetry and Leibniz rule are automatically satisfied for $\{\cdot,\cdot\}_\lambda$, the only remaining condition is the Jacobi identity. There are other equivalent forms for the compatibility condition. For example, let us set for $F \in C^\infty(\mathcal{M})$, $X_F \stackrel{def}{=} P_0 dF$ and $Y_F \stackrel{def}{=} P_1 dF$. It is not hard to show that the compatibility condition can be rewritten as

$$L_{X_F} P_1 + L_{Y_F} P_0 = 0 \qquad \forall F \in C^\infty(\mathcal{M}),$$

where $L_X$ denotes the Lie derivative along the vector field $X$.

**Definition 1.3.** *Let $(\mathcal{M}, P_0, P_1)$ be a bi-Hamiltonian manifold. A vector field $X$ on $\mathcal{M}$ is said to be a* bi-Hamiltonian vector field *if it is Hamiltonian with respect to both $P_0$ and $P_1$, that is, if there exist two functions $H_0$ and $H_1$ such that*

(1.7) $$X = P_0 dH_1 = P_1 dH_0.$$

The next definition is the essential component behind Lenard's construction. The latter is widely used to show that certain Hamiltonian models, such as the KdV, possess infinitely many conserved quantities that are in involution with one another.

**Definition 1.4.** *Let $(\mathcal{M}, P_0, P_1)$ be a bi-Hamiltonian manifold. A bi-Hamiltonian hierarchy is a sequence $\{H_k\}_{k \geq 0}$ of functions on $\mathcal{M}$ fulfilling* Lenard's recursion relations

$$\{\cdot, H_{k+1}\}_0 = \{\cdot, H_k\}_1, \qquad k \geq 0, \tag{1.8}$$

*and the additional condition $\{\cdot, H_0\}_0 = 0$. In terms of Poisson tensors: $P_0 dH_{k+1} = P_1 dH_k$, with $H_0$ a Casimir function of $P_0$: $P_0 \, dH_0 = 0$.*

A bi-Hamiltonian hierarchy immediately gives rise to an infinite sequence of bi-Hamiltonian vector fields

$$X_k = P_0 dH_k = P_1 dH_{k-1}. \tag{1.9}$$

They all commute, since the functions of a bi-Hamiltonian hierarchy are in involution with respect to both Poisson brackets.

**Remark 1.5.** Let $\{H_k\}$ be a bi-Hamiltonian hierarchy. Let us introduce the function

$$H(\lambda) = \sum_{k \geq 0} H_k \lambda^{-k}; \tag{1.10}$$

then, as an immediate consequence of the definition of bi-Hamiltonian hierarchy, we have that $H(\lambda)$ is a Casimir function of the Poisson pencil $\{\cdot, \cdot\}_\lambda$. Vice-versa, let $H(\lambda)$ be a Casimir function of $\{\cdot, \cdot\}_\lambda$ which can be developed in a Laurent series: $H(\lambda) = \sum_{k \geq -n} H_k \lambda^{-k}$, with a suitable $n$; then the coefficients $\{H_k\}$ form a bi-Hamiltonian hierarchy.

From the above simple remark we get a method for seeking bi-Hamiltonian hierarchies, namely: seek the Casimir functions of the Poisson pencil, which are deformations of Casimir functions of $P_0$. This method has displayed its effectiveness in the case of a number of integrable models, such as the KdV hierarchy, the Toda lattice hierarchy, etc. See [12, 13, 19, 22].

**Remark 1.6.** Let $\{H_k\}$ be a bi-Hamiltonian hierarchy, and $X_k = P_0 dH_k = P_1 dH_{k-1}$ the corresponding bi-Hamiltonian vector fields; let us introduce the functions

$$H^k(\lambda) \stackrel{\text{def}}{=} \sum_{j=0}^{k} H_j \lambda^{k-j}. \tag{1.11}$$

Then it is easy to show that, for all $\lambda$, the vector fields $X_k$ are Hamiltonian with respect to $P_\lambda$, with Hamiltonian $H^{k-1}(\lambda)$:

$$X_k = P_\lambda dH^{k-1}(\lambda). \tag{1.12}$$

Therefore, the vector fields $X_k$ are not only *bi*Hamiltonian, but they are Hamiltonian with respect to any bracket of the Poisson pencil.

Let us briefly recall now the construction of the KdV and KP hierarchies of the bi-Hamiltonian point of view. For more details the reader is referred to [12, 13, 18].

Let $\mathcal{U}$ be the infinite dimensional manifold of scalar valued $C^\infty$–functions on $S^1$, endowed with the Poisson brackets

$$\{F,G\}_0 = 2\int_{S^1} dF(dG)_x\, dx \tag{1.13}$$

$$\{F,G\}_1 = \frac{1}{2}\int_{S^1} (dF(dG)_{xxx} - 4u(dG)_x - 2u_x dG)\, dx, \tag{1.14}$$

so that the Poisson pencil is

$$(P_\lambda)_u v = -\frac{1}{2}v_{xxx} + 2(u+\lambda)v_x + u_x v. \tag{1.15}$$

In these formulas, $u \in \mathcal{U}$ is the point, and $v$ is a co-vector at $u$, the pairing between $v$ and a generic tangent vector $\dot{u}$ being given by

$$\langle v, \dot{u}\rangle = \int_{S^1} v(x)\dot{u}(x)\, dx. \tag{1.16}$$

In order to show that the Poisson pencil (1.15) admits a Casimir function we have to find an exact 1-form $v(\lambda) = dH(\lambda)$ on $\mathcal{U}$, such that

$$P_\lambda v(\lambda) = -\frac{1}{2}v(\lambda)_{xxx} + 2(u+\lambda)v(\lambda)_x + u_x v(\lambda) = 0. \tag{1.17}$$

Let $v(\lambda)$ be the unique solution of the equation

$$\frac{1}{4}v_x^2 - \frac{1}{2}vv_{xx} + (u+\lambda)v^2 = \lambda \tag{1.18}$$

of the form $v(\lambda) = 1 + \sum_{i\geq 1} v_i \lambda^{-i}$. Then it can be shown that $v(\lambda)$ is an exact 1–form. Indeed, if we set $\lambda = z^2$ and

$$h(z) = \frac{z}{v} + \frac{v_x}{2v}, \tag{1.19}$$

so that

$$h_x + h^2 = u + z^2, \tag{1.20}$$

we have that the 1–form $v$ is the differential of the function

$$H = 2z\int_{S^1} h\, dx \tag{1.21}$$

which, therefore, is a Casimir of the Poisson pencil. Thus the Casimir functions of the Poisson pencil (1.15) can be computed by solving the Riccati equation (1.20).

The basic object in the previous discussion is the map $J$ from the manifold $\mathcal{U}$ to the space of Laurent series assigning to the point $u \in \mathcal{U}$ the corresponding Hamiltonian density $h(u)$, i.e., the solution of the Riccati equation (1.20). Since $J$ allows us to compute the Hamiltonians, we call it the *momentum map* associated with the Abelian algebra acting on $\mathcal{U}$ by means of the KdV hierarchy. As a consequence of the involutivity relations between the functions $H_j$, when $u$ evolves according to the KdV flows $h(u)$ must obey conservation laws of the form

$$\frac{\partial h}{\partial t_j} = \partial_x H^{(j)}, \quad j = 1, 2, \cdots \tag{1.22}$$

where the $H^{(j)}$ are suitable current densities. The important point is that one can write the $H^{(j)}$ directly in terms of a generic Laurent series of the form

$$h(z) = z + \sum_{j \geq 1} h_j z^{-j}. \tag{1.23}$$

Then the equations (1.22) become an infinite systems of partial differential equations in the $\{h_j\}_{j \geq 1}$, which is equivalent to the KP hierarchies. These ideas will be implemented in detail for the case of the HD equation in the next section.

## 2. The Harry Dym Hierarchy and its bi-Hamiltonian Structure

In this paragraph we start studying the HD hierarchy from the bi-Hamiltonian view point in analogy with the scheme developed earlier for KdV equation. Again, let $\mathcal{U}$ be the space of scalar valued $C^\infty$–functions on $S^1$, endowed with the Poisson brackets [24, 25]

$$\{F, G\}_0 = \frac{1}{2} \int_{S^1} dF (dG)_{xxx} \, dx \tag{2.1}$$

$$\{F, G\}_1 = - \int_{S^1} (2u(dG)_x + u_x dG) \, dx. \tag{2.2}$$

The corresponding Poisson tensors are given by

$$(P_0)_u = -\frac{1}{2} \partial_{xxx} \tag{2.3}$$

$$(P_1)_u = -2u \partial_x - u_x, \tag{2.4}$$

so that the Poisson pencil is

$$(P_\lambda)_u v = -\frac{1}{2} v_{xxx} + \lambda (2u v_x + u_x v). \tag{2.5}$$

It is straight-forward to show that $P_\lambda$ is a Poisson tensor, so that the above defined brackets are compatible and $\mathcal{U}$ is a bi-Hamiltonian manifold.

Our immediate goal is to show that, as in the KdV case, the Poisson pencil (2.5) admits a Casimir function in the form of a Laurent series. This allows us to write a hierarchy of integrable vector fields for which HD is one of them. The important point is that the search for a Casimir function for (2.5) is also related to solving a Riccati equation. To wit, we start by finding an exact 1-form $w(\lambda) = dK(\lambda)$ on $\mathcal{U}$, such that

$$P_\lambda w(\lambda) = -\frac{1}{2} w(\lambda)_{xxx} + \lambda (2u w(\lambda)_x + u_x w(\lambda)) = 0. \tag{2.6}$$

We integrate once this equation noticing that

$$w \left( -\frac{1}{2} w_{xxx} + \lambda (2u w_x + u_x w) \right) = \frac{d}{dx} \left( \frac{1}{4} w_x^2 - \frac{1}{2} w w_{xx} + \lambda u w^2 \right), \tag{2.7}$$

and therefore we consider the equation

$$\frac{1}{4} w_x^2 - \frac{1}{2} w w_{xx} + \lambda u w^2 = \lambda. \tag{2.8}$$

This equation admits a unique solution of the form $w(\lambda) = w_0 + \sum_{i \geq 1} w_i \lambda^{-i}$, and the coefficients $w_i$ can be computed iteratively. In order to show that $w(\lambda)$ is an exact 1–form, we set $\lambda = z^2$, and remark that equation (2.8) can be written in the form of a *Riccati equation*

$$\tag{2.9} \left(\frac{z}{w} + \frac{w_x}{2w}\right)_x + \left(\frac{z}{w} + \frac{w_x}{2w}\right)^2 = uz^2.$$

For this reason, we set

$$\tag{2.10} k(z) = \frac{z}{w} + \frac{w_x}{2w},$$

so that (2.9) takes the form

$$\tag{2.11} k_x + k^2 = uz^2.$$

Then we consider a curve $u(t)$ in $\mathcal{U}$, and denote by $w(t)$ the solution of (2.8) at the points of this curve. By differentiating equation (2.11) with respect to a time $t$, we get $z^2 \dot{u} = \dot{k}_x + 2k\dot{k}$ and therefore

$$\tag{2.12} \langle w, \dot{u} \rangle = \frac{1}{z^2} \int_{S^1} w(x)(\dot{k}_x + 2k\dot{k})\, dx = \frac{2}{z^2} \int_{S^1} (-\frac{1}{2}w_x + kw)\dot{k}\, dx = \frac{d}{dt}\left(\frac{2}{z}\int_{S^1} k\, dx\right),$$

where we used equation (2.10) in the form

$$\tag{2.13} -\frac{1}{2}w_x + kw = z.$$

Equation (2.12) shows that the 1–form $w$ is the differential of the function

$$\tag{2.14} K = \frac{2}{z} \int_{S^1} k\, dx$$

which, therefore, is a Casimir of the Poisson pencil.

Thus the Casimir functions of the Poisson pencil (2.5) can be computed by solving the Riccati equation (2.11); and note that its unique solution has the asymptotic expansion in powers of $z$,

$$\tag{2.15} k(z) = k_{-1}z + \sum_{j \geq 0} k_j z^{-j}.$$

It can be shown that $k_{2j}$ is a total derivative for all $j$, so that $K$ is a Laurent series in $\lambda$. For their importance, we summarize these facts in the following

**Proposition 2.1.** *Let $k$ be the unique solution of the Riccati equation (2.11)*

$$k_x + k^2 = uz^2$$

*admitting the asymptotic expansion (2.15). Then, its integral $K(\lambda)$ given by (2.14) is a Casimir function of the Poisson pencil (2.5).*

Now we have an algebraic technique to compute the Casimir function. Before proceeding, we give an example how it works for the first orders of $z$. First, write the Riccati equation (2.11)

$$
\begin{aligned}
(z^2) & & k_{-1}^2 &= u \\
(z) & & k_{-1,x} + 2k_0 k_{-1} &= 0 \\
(1) & & k_{0,x} + k_0^2 + 2k_{-1}k_1 &= 0 \\
(z^{-1}) & & k_{1,x} + 2k_0 k_1 + 2k_{-1}k_2 &= 0 \\
(z^{-2}) & & k_{2,x} + k_1^2 + 2k_0 k_2 + 2k_{-1}k_3 &= 0 \\
& \vdots & & \vdots
\end{aligned}
\tag{2.16}
$$

for the first order coefficients. Thus we can solve (2.16) recursively,

$$k_{-1} = \sqrt{u} \tag{2.17}$$

$$k_0 = -\frac{1}{4}\frac{u_x}{u} \tag{2.18}$$

$$k_1 = -\frac{5}{32}\frac{u_x^2}{u^2\sqrt{u}} + \frac{1}{8}\frac{u_{xx}}{u\sqrt{u}} \tag{2.19}$$

$$\vdots \tag{2.20}$$

At this point we have the expression in $u$ and its derivatives of the coefficients $k_j$ of $k(z)$. Following the previous proposition, we can construct the Casimir $K(\lambda) = 2z^{-1}\int_{S^1} k\, dx$ of $P_\lambda$. Hence, $K = K_0 + K_2\lambda^{-1} + K_4\lambda^{-2} + \ldots$, whenever

$$K_j = 2\int_{S^1} k_{j-1}\,.$$

The functions $K_{2j}$ form a bi-Hamiltonian hierarchy and the vector fields of the hierarchy are now defined by $X_{2j+1} = P_0 dK_{2j} = P_1 dK_{2j-2}$. A simple calculation shows that the vector field corresponding to the Harry Dym equation (1.1) after the change $u = 1+q$ and a simple time rescaling is precisely

$$X_3(u) = -\frac{1}{2}\left(\frac{1}{\sqrt{u}}\right)_{xxx}\,.$$

Finally, we conclude this section observing that all the HD equations are Hamiltonian not only with respect to both Poisson tensors $P_0$ and $P_1$, but also with respect to the Poisson pencil $P_\lambda$. Indeed, from a simple computation (see also Remark 1.6), we have that

$$X_{2j+1} = P_\lambda dK^{(2j-2)}(\lambda)\,, \tag{2.21}$$

where [2] $K^{(2j-2)}(\lambda) = (z^{2j-2}K(\lambda))_+$.

---

[2] Here we denote by the $(S(z))_+$ the non-negative powers of a Laurent expansion $S(z)$ in $z$.

## 3. The Harry Dym-KP hierarchy

The flows defined by equation (2.21) are naturally plunged into a hierarchy of commuting flows with an infinite number of fields in a similar way that the KdV is plunged into the KP hierarchy. In fact, Carillo and Oevel [11] have obtained such Harry Dym-KP hierarchy with the help of pseudo-differential operators. We close this section explaining how to relate the Harry Dym-KP hierarchy constructed in [11] with the one constructed here following the same techniques explained in [18].

One of the key remarks of the present note is that we can construct such hierarchy "from first principles" as the one that governs the evolution of $k(u)$, solution of (2.11), as $u$ evolves according to the Harry Dym hierarchy. The basic object is once again the map from the manifold $\mathcal{U}$ to the algebra of Laurent series, defined by the Riccati equation (2.11). Since a solution $k(z)$ of it allows us to compute the Hamiltonians, we call it the *momentum map* associated with the Abelian algebra acting on $\mathcal{U}$ by means of the HD hierarchy. As a consequence of the involutivity relations between the functions $K_j$, when $u$ evolves according to the HD flows, $k(u)$ must obey conservation laws of the form

$$\frac{\partial k}{\partial t_{2j+1}} = \partial_x K^{(2j+1)},$$

for some suitable current densities $K^{(2j+1)}$. Our problem is to find a nice expression for the currents $K^{(j)}$, which can be computed following the ideas developed in [13] for KdV. From equation (2.11) we have that

$$(3.1) \qquad (\partial_x + 2k)\left(\frac{\partial k}{\partial t_j}\right) = z^2 \frac{\partial u}{\partial t_j},$$

while the HD equations (2.21) can be written as

$$(3.2) \qquad \frac{\partial u}{\partial t_{2j+1}} = (\partial_x + 2k)(\tfrac{1}{2}\partial_x)(-\partial_x + 2k)(\lambda^j w)_+,$$

since

$$(3.3) \qquad -\frac{1}{2}w_{xxx} + \lambda u w_x + u_x w = (\partial_x + 2k)(\tfrac{1}{2}\partial_x)(-\partial_x + 2k) \cdot v.$$

Therefore

$$(3.4) \qquad \begin{aligned} (\partial_x + 2k)\left(\tfrac{\partial k}{\partial t_{2j+1}}\right) &= z^2(\partial_x + 2k)(\tfrac{1}{2}\partial_x)\left(-(z^{2j-1}w)_{+,x} + 2k(z^{2j-1}w)_+\right) \\ &= (\partial_x + 2k)\partial_x z^2 \left(-\tfrac{1}{2}(z^{2j-1}v)_{+,x} + h(z^{2j-1}v)_+\right), \end{aligned}$$

which implies

$$(3.5) \qquad K^{(2j+1)} = -\frac{1}{2}(\lambda^j v)_{+,x} + h(\lambda^j v)_+.$$

This is the formula for the current densities that appear in our local conservation laws. Observe that we can put $K^{(2j)} = z^{2j}$, since the even times are trivial, and we extend the definition of the currents for every $j$.

The fundamental observation allowing to generalize this procedure to the HD-KP theory is that we can compute the currents $K^{(j)}$ directly from $k$, without computing the series $w(\lambda)$, in perfect analogy with the KdV theory [13].

A key object in our construction is the notion of *Faà di Bruno polynomials*. They are defined [23] as $k^{(j)} = (\partial_x + k)^j \cdot 1$, for $j \geq 0$.

**Proposition 3.1.** *The above defined currents $K^{(l)}$, for $l \geq 2$, are determined directly from $k$, and in particular they satisfy*

1. $K^{(l)} = z^l + O(z)$
2. $K^{(l)} \in K''_+$, where $K''_+ \stackrel{def}{=} \langle k^{(2)}, k^{(3)}, \dots \rangle$ *(linear span over $C^\infty$–functions)*.

*Sketch of the proof:* For $l$ even, there is nothing to show. To show the first item above for $l = 2j+1$ we write $(z^{2j-2}w)_+ = z^{2j-2}w - (z^{2j-2}w)_-$ and make use of the relation $-\frac{1}{2}w_x + kw = z$. As for the second item, we recall that $(z^{2j-2}w)_+ = \sum_{k=0}^{j-1} w_{2k} z^{2(j-k-1)}$ and thus

$$K^{(2j+1)} = -\frac{1}{2} \sum_{k=0}^{j} w_{2k,x} z^{2(j-k)} + \sum_{k=0}^{j} w_{2k} z^{2(j-k)} k \ .$$

To conclude the proof of the second item, it remains to show that $z^{2k} k^{(l)} \in K''_+$ and hence

$$z^2 k^{(l)} = z^2 (\partial_x + k)^l (1) = (\partial_x + k)^l (z^2) \in K''_+$$

since $(\partial_x + k) K''_+ \subset K''_+$. Thus, for example,

$$z^4 = z^2 z^2 = z^2 \frac{1}{u} k^{(2)} = \frac{1}{u} z^2 k^{(2)} \in K''_+ \ ,$$

and so on. Q. E. D.

The previous proposition is the central point to formulate the KP theory for the HD equations. It means that we can compute the currents $K^{(j)}$, with $j \geq 2$, starting with any Laurent series $k(z)$ of the form

(3.6) $$k = k_{-1} z + \sum_{l \geq 0} k_l z^{-l}$$

by following the two requirements in items (1) and (2) of Proposition 3.1 and not necessarily those satisfying (2.11). For example, it is possible to write

(3.7) $$K^{(l)} = c_l k^{(l)} + \cdots + c_2 k^{(2)} \ ,$$

with $c_j$ coefficients, that depend on the coefficients of $k$, determined by the asymptotic expansion. At this point the local conservation laws became evolution equations in the space of Laurent series of the form (3.6) and live independently of the HD equations.

**Definition 3.2.** *The Harry Dym-Kadomtsev-Petviashvili (HD-KP) hierarchy is the system of vector fields acting on the space of Laurent series of the form (3.6) defined by*

$$\frac{\partial k}{\partial t_l} = \partial_x K^{(l)} \quad , \ l = 2, 3, \cdots$$

where the conserved densities $K^{(l)}$ are defined by the prescription of items (1) and (2) of Proposition 3.1.

To obtain the HD hierarchy defined above all that we need to do is to impose the constraint

$$K^{(2)} = z^2. \tag{3.8}$$

Indeed, it is easy to see that (3.8) is equivalent to the Riccati equation for Harry Dym: express, according to (3.7), the current $K^{(2)}$ in the basis of Faà di Bruno polynomials, i.e.,

$$K^{(2)} = (k_{-1})^{-2} k^{(2)} \; ;$$

then (3.8) becomes the Riccati equation (2.11) if we put $u = k_{-1}{}^2$. This implies that $K^{(2k)} = z^{2k}$ for $k = 1, 2, \cdots$, and thus all the even flows are stationary.

In [11] a hierarchy of commuting flows was constructed, making use of the notion of pseudo-differential operators, that includes the Harry Dym flow as one of its reductions. To do that, they consider the operator

$$L_{\text{Dym}} \stackrel{\text{def}}{=} q_{-1}\partial_x + q_0 + w_1 \partial_x^{-1} + q_2 \partial_x^{-2} + \cdots . \tag{3.9}$$

The hierarchy is then defined in Lax form as

$$\frac{\partial L_{\text{Dym}}}{\partial t_j} = \left[ \left(L_{\text{Dym}}^j\right)_{\geq 2}, L_{\text{Dym}} \right] \, , j \geq 2 \tag{3.10}$$

where $(S)_{\geq 2}$ means the projection on the differential operators of order $\geq 2$ of the pseudo-differential operator $S$.

This definition is equivalent to ours, after a change of variable which follows the one in [13]. Start by defining the negative Faà di Bruno polynomials $k^{(j)}$, with $j < 0$, which can be obtained from $k^{(0)} = 1$ by solving backwards the equations

$$k^{(j+1)} = k_x^{(j)} + k k^{(j)},$$

and expand $z$ on the basis $\{k^{(j)}\}_{j \in Z}$:

$$z = q_{-1} k + \sum_{j \geq 0} q_j k^{(-j)}.$$

This establishes a change of variables between $\{k_j\}_{j \geq -1}$ and $\{q_j\}_{j \geq -1}$. Then it can be shown that equations (3.10) coincide with the HD-KP hierarchy given in Definition 3.2.

## 4. The Evolution of the Currents

We shall now study in more detail the evolution of the currents $K^{(j)}$ when $k$ evolves according to the HD-KP hierarchy. Our first remark is that we can interpret the HD-KP equations as the commutativity conditions

$$\left[\partial_{t_j} + K^{(j)}, \partial_x + k\right] = 0$$

of the operators $\partial_{t_j} + K^{(j)}$ and $\partial_x + k$. Since $K^{(j)} \in K''_+$ and $K''_+$ is invariant with respect to $\partial_x + k$, it follows that

$$(4.1) \qquad \left(\partial_{t_j} + K^{(j)}\right) k^{(l)} = \left(\partial_{t_j} + K^{(j)}\right)(\partial_x + k)^l 1 = (\partial_x + k)^l K^{(j)} \in K''_+,$$

for every $l \geq 0$ and $j \geq 2$. Therefore,

$$(4.2) \qquad \left(\partial_{t_j} + K^{(j)}\right)(K^{(i)}) \in K''_+ .$$

As in the KdV case, we observe that the currents $K^{(j)}$ form a basis of $K''_+$. This fact allows us to look at the equations (4.2) as a combination of the currents with suitable coefficients. In more detail, if we write

$$(4.3) \qquad K^{(i)} = z^i + K^i_{-1} z + \sum_{l \geq 0} K^i_l z^{-l} , \qquad i \geq 2 ,$$

then, equations (4.2) can be written in the explicit form

$$
\begin{aligned}
\frac{\partial K^{(i)}}{\partial t_j} + K^{(i)} K^{(j)} &= K^{(i+j)} + K^i_{-1} K^{(j+1)} + K^j_{-1} K^{(i+1)} \\
&\quad + K^i_{-1} K^j_{-1} K^{(2)} + K^i_0 K^{(j)} + K^j_0 K^{(i)} \\
&\quad + \sum_{l=1}^{i-2} K^j_l K^{(i-l)} + \sum_{l=1}^{j-2} K^i_l K^{(j-l)}, \quad \text{for } i,j \geq 2,
\end{aligned}
$$
(4.4)

which gives the evolution of the currents. A similar expression can be found for the evolution of $k$. Indeed, (4.1) entails that

$$\frac{\partial k}{\partial t_j} + k K^{(j)} \in K''_+ ,$$

so that

$$
\begin{aligned}
\frac{\partial k}{\partial t_j} + k K^{(j)} &= k_{-1} K^{(j+1)} + k_0 K^{(j)} \\
&\quad + K^j_{-1} k_{-1} K^{(2)} + \sum_{l=1}^{j-2} k_l K^{(j-l)}, \quad \text{for } j \geq 2.
\end{aligned}
$$
(4.5)

Of course, this is not different from the definition of the HD-KP hierarchy, but it does not explicitly contain $x$-derivatives. Furthermore, the following equations also hold

$$
\begin{aligned}
\partial_x K^{(j)} + k K^{(j)} &= k_{-1} K^{(j+1)} + k_0 K^{(j)} \\
&\quad + K^j_{-1} k_{-1} K^{(2)} + \sum_{l=1}^{j-2} k_l K^{(j-l)}, \quad \text{for } j \geq 2,
\end{aligned}
$$
(4.6)

$$(4.7) \qquad \partial_x k + k^2 = k^2_{-1} K^{(2)} ,$$

coming from the fact that the operator $\partial_x + k$ sends the span $K'_+ \stackrel{\text{def}}{=} \langle k^{(j)} \rangle_{j \geq 1}$ into $K''_+$.

In the next section we shall see that, in analogy with the Central-System equations for the KP hierarchy, equations (4.4), (4.6), (4.5), and (4.7) will give rise to the *Central-System equations for the Harry Dym hierarchy.*

We stress that the above equations (4.4–4.7) underscore the fact that the evolution of the Laurent coefficients of the currents $K^{(i)}$ and $k$ undergoes a *dynamical system* which does not involve spatial derivatives. We also remark that such system is different from the one obtained in [21] since here $k$ is not monic anymore.

## 5. The Central System of the Harry Dym Hierarchy

In the previous section we analyzed the evolution of the currents when the HD-KP equations hold. We notice that all the coefficients in the equations (4.4–4.7) are dependent on the coefficients $k_j$ of $k$ through the Faà di Bruno polynomials as in Proposition 3.1. The main point to pass to the Central System is to consider the currents (4.3) as arbitrary elements of $K''_+$ independent of $k$. This is one step further to considering $k$ independent of the Riccati equation when we passed from the local conservation laws to the HD-KP equations. In this new picture, the series $k$ does not play any special role, and we rename it by

$$K^{(1)} = K^1_{-1} z + \sum_{l \geq 0} K^1_l z^{-l}.$$

We have shown that the evolution of the currents is given by the equations

(5.1) $$\left(\partial_{t_j} + K^{(j)}\right) K^{(l)} \in K''_+, \quad \text{for } l, j \geq 1.$$

It is not hard to see that they can be rewritten in the form

(5.2) $$\partial_{t_j} K^{(l)} = -\pi''_-\left(K^{(j)} K^{(l)}\right), \quad \text{for } l, j \geq 1,$$

where we denote by $\pi''_-$ and $\pi''_+$ the projections of the space $L$ of all Laurent series onto the subspaces $K''_+$ and $K''_-$ respectively, defined by the following decomposition

(5.3) $$L = \langle K^{(2)}, K^{(3)}, \dots \rangle \oplus \langle z, 1, z^{(-1)}, \dots \rangle = K''_+ \oplus K''_-.$$

We call (5.2) the *HD-Central System* (HD-CS) associated to the HD-KP theory. We conclude the present section by analyzing a few of its properties.

The first remark is that the evolution of the currents given by the Central-System equations for the Harry Dym hierarchy form a system of nonlinear ordinary differential equations of Riccati type in the components of the fields $K^{(i)}$, for $i \geq 1$. The second remark is that the flows defined above commute with one another. To show that, we start with the remark that along the HD-CS flows (5.2), we have that

$$\frac{\partial K^{(j)}}{\partial t_i} = \frac{\partial K^{(i)}}{\partial t_j}.$$

Then, we compute the commutator of two vector fields of the HD-CS hierarchy on a generic current $K^{(i)}$. Let us call such fields say $\mathbf{X}_j$ and $\mathbf{X}_l$, and notice that

$$[\mathbf{X}_j, \mathbf{X}_l]\left(K^{(i)}\right) = \frac{\partial}{\partial t_j}\frac{\partial K^{(i)}}{\partial t_l} - \frac{\partial}{\partial t_l}\frac{\partial K^{(i)}}{\partial t_j} = \left[\frac{\partial}{\partial t_j} + K^{(j)}, \frac{\partial}{\partial t_l} + K^{(l)}\right] K^{(i)}.$$

Because of the specific form of the Laurent series of $K^{(i)}$ the above commutator belongs to $K''_-$. On the other hand, because of the invariance condition (5.2) that defines the flows, we have that the commutator belongs to $K''_+$, and hence vanishes.

As a consequence we can apply the *spatialization process*, already described in earlier works [21]. This procedure is based on the following:

**Remark 5.1.** *Assume we have a family $\{\mathbf{X}_j\}$ of commuting vector fields on a manifold. Then, two reductions processes are natural. The first one is the projection along the integral curves of one of the vector fields. The second one, is the restriction to the set of zeroes of a fixed vector field.*

It easy to see that, in particular, by setting $t_1 = x$ and using the first flow to write all the equations in terms of $k \stackrel{\text{def}}{=} K^{(1)}$ and its derivatives with respect to $x$ we obtain the HD-KP hierarchy. This corresponds to the first reduction mentioned above. The second step is to consider the stationary flows with respect to the $t_2$ flow given by the constraint $K^{(2)} = z^2$. This gives back the Riccati equation and the HD hierarchy we saw in Section 2.

## 6. Concluding Remarks

In the present work we have carried out for the Harry Dym equation, the first steps of the program that was developed in [13, 18] for the KdV and KP hierarchies. Namely, we constructed the HD hierarchy, the HD-KP hierarchy and the corresponding Central System. The program in [13, 18] lead to the linearization of the flows in the KP hierarchy and the construction of polynomial $\tau$-functions [17, 19]. This was achieved by using the notion of Darboux coverings and introducing the *modified Central System*. This would be a natural follow up of the present work.

## Acknowledgments

We thank Franco Magri for stimulating discussions and suggestions. M. Pedroni was supported by MURST under grant "Geometry of Integrable Systems." V. Sciacca was supported by MURST under grant "Non linear Mathematical Problems of Wave Propagation and Stability in Models of Continuous Media." J. P. Zubelli was supported by the fellowship from the Brazilian National Science Research Council (CNPq) of number 200258/84-2 ("Bolsista do CNPq - Brasil"), a grant from the State of Rio de Janeiro Research Foundation (FAPERJ) number 170.369/2001, and the research at MSRI was supported in part by NSF grant DMS-9810361.

Marco Pedroni, Dipartimento di Matematica, Università di Genova, Via Dodecaneso 35, I-16146, Genova, Italy
  *E-mail address*: `pedroni@dima.unige.it`

Vincenzo Sciacca, Dipartimento di Matematica e Applicazioni, Università di Palermo, Via Archirafi 34 - CAP 90123, Palermo, Italy
  *E-mail address*: `sciacca@dipmat.math.unipa.it`

Jorge P. Zubelli, IMPA, Est. D. Castorina 110, RJ 22460-320, Rio de Janeiro, Brazil
  *E-mail address*: `zubelli@impa.br`